# Leveraging GPT-4 for Vulnerability-Witnessing Unit Test Generation

Gábor Antal
gabor.antal@frontendart.com
antal@inf.u-szeged.hu
FrontEndART Software Ltd.
University of Szeged
Szeged, Hungary

Dénes Bán
zealot@inf.u-szeged.hu
University of Szeged
Szeged, Hungary

Martin Isztin
isztin@inf.u-szeged.hu
University of Szeged
Szeged, Hungary

Rudolf Ferenc
ferenc@inf.u-szeged.hu
University of Szeged
Szeged, Hungary

Péter Hegedűs
peter.hegedus@frontendart.com
hpeter@inf.u-szeged.hu
FrontEndART Software Ltd.
University of Szeged
Szeged, Hungary

## Abstract

In the life-cycle of software development, testing plays a crucial role in quality assurance. Proper testing not only increases code coverage and prevents regressions but it can also ensure that any potential vulnerabilities in the software are identified and effectively fixed. However, creating such tests is a complex, resource-consuming manual process. To help developers and security experts, this paper explores the automatic unit test generation capability of one of the most widely used large language models, GPT-4, from the perspective of vulnerabilities. We examine a subset of the VUL4J dataset containing real vulnerabilities and their corresponding fixes to determine whether GPT-4 can generate syntactically and/or semantically correct unit tests based on the code before and after the fixes as evidence of vulnerability mitigation. We focus on the impact of code contexts, the effectiveness of GPT-4's self-correction ability, and the subjective usability of the generated test cases. Our results indicate that GPT-4 can generate syntactically correct test cases 66.5% of the time without domain-specific pre-training. Although the semantic correctness of the fixes could be automatically validated in only 7. 5% of the cases, our subjective evaluation shows that GPT-4 generally produces test templates that can be further developed into fully functional vulnerability-witnessing tests with relatively minimal manual effort.

Therefore, despite the limited data, our initial findings suggest that GPT-4 can be effectively used in the generation of vulnerability-witnessing tests. It may not operate entirely autonomously, but it certainly plays a significant role in a partially automated process.

## 1 Introduction

To this day, many software releases contain bugs and vulnerabilities. Security issues that might make it to countless production servers or consumer machines remain unnoticed until someone exploits them. Unfortunately, even an experienced software engineer cannot ensure the reliability of a program code just by visual inspection, as a piece of code might be harmless in one environment but vulnerable in another. Fixing such flaws requires significant time, effort and financial resources. And even though these vulnerabilities can be identified and mitigated with adequate testing, writing the tests for them is itself a resource-intensive task.

To avoid the need for individual discovery, databases exist that document commonly disclosed vulnerabilities. One of these is the CVE (*Common Vulnerabilities and Exposures*) [6] database, containing information about cybersecurity vulnerabilities in real systems. Each entry has an identifier representing one specific vulnerability. A more general categorization of these vulnerabilities is done using CWE (*Common Weakness Enumeration*)[7] identifiers. Adapting important domain knowledge from these databases would be a valuable addition to software testing, but it still requires significant manual effort to do so. From an automation perspective, Large Language Models (LLMs) have become widespread as productivity-enhancing technologies, and due to the opportunities they offer, they are also used in various areas, such as software development [12, 17]. Numerous studies report on the automatic program-fixing capabilities of this emerging technology [3, 8, 15, 25, 27], but there is little information on generating unit tests to challenge these fixes – or, indeed, any other manually created fixes for potential vulnerabilities.

This paper aims to generate unit tests for recognized vulnerabilities in practice. Our goal, therefore, is not finding or even fixing previously unknown vulnerabilities, but generating proofs (so called vulnerability-witnessing test cases) for existing ones. The benefits of being able to complement such vulnerability-fix pairs with corresponding tests would be two-fold. The **developer community** could leverage this approach to semi-automatically generate regression tests for vulnerabilities they naturally encounter in their own code. The **research community** could utilize it to annotate existing but untested vulnerabilities, thereby expanding the datasets needed for benchmarking and evaluating new models.

For the evaluation, we used a subset of the VUL4J [4] dataset, which contains information about vulnerabilities in real-world systems written in Java. The dataset includes the vulnerability fixes, as well as the corresponding CVE (and often even the CWE) identifiers. It has additional features, such as the ability to easily switch between vulnerable and fixed versions of the projects with the provided environment, access to the changes made for fixing the vulnerabilities, and the original unit tests used in the project.

Our evaluation subset contains 50 entries selected to cover as many types of vulnerabilities as possible. Each entry was divided into focal contexts [22] consisting of four levels (referred to as L0-L3). We then used one of the most well-known language models, OpenAI's generative pre-trained transformer model, GPT, specifically the GPT-4 Turbo version to process these contexts. For the model's input text (i.e., the prompt), we provided the appropriate source code segments from before and after the fix, plus a number of other, potentially useful modifiers including:

- A **role** allowing the LLM to give more accurate and contextually appropriate responses; in our case, the model was asked to act as a *senior software tester*,
- **Emotional stimulation**, which has been shown to be able to influence the accuracy of an LLM's response, and
- The relevant **CWE** identifier to check whether it helps generate better test cases (see Section 5.2).

After using an automated environment to produce the context and the prompt, which was then passed along to the LLM, the classification of the results was divided into two parts. First, our framework processed and evaluated the response it received from the LLM in an automated fashion. Second, we also manually reviewed the subjective usability of the generated code.

The research aims to answer the following questions:

- **RQ1**: What percentage of the generated unit tests code is syntactically correct?
- **RQ2**: What percentage of the generated unit tests code is semantically correct?
- **RQ3**: Based on the subjective evaluation, how useful are the generated unit tests?
- **RQ4**: How does the context affect the accuracy of the generation?

Our results indicate that GPT-4 can generate syntactically correct test cases 66.5% of the time without any domain-specific pre-training, which is encouraging considering that no technical background information was provided during prompting. Based on the actual unit test execution, the generated test cases were also semantically correct in 7.5% of the cases. However, during our subjective, manual evaluation, we found that GPT-4 generated useful templates for developers in 68.5% of cases. This demonstrates that, with sufficient fine-tuning (either by manually modifying the source code or presumably using a more refined prompt), large language models can be beneficial in a (semi-)automated workflow.

The structure of the paper is as follows: Section 2 discusses related work, followed by the research methodology in Section 3. Section 4 presents our results in detail, Section 5 discusses special cases and additional observations. Section 6 covers the limitations and future possibilities, and finally, Section 7 concludes the paper.

## 2 Related Work

### 2.1 LLMs in vulnerability analysis

LLMs are already widely used in vulnerability analysis. One of the potential directions is vulnerability detection where, according to Zhou et al. [28], GPT-3.5 matches, while GPT-4 outperforms state-of-the-art detection methods. One step beyond pure detection is *automated program repair* (APR), which falls closer to our use-case of code generation, and is the subject of numerous studies. Zhang et al. [26] compared the APR capabilities of the VRepair transfer-learning neural network model with several large language models, including CodeBERT, UniXcoder, and CodeGPT. Their main result revealed that large language models performed repairs with an accuracy between 10.21% and 22.23%. In another study, Fu et al. [9] specifically used ChatGPT to evaluate the model's APR capabilities on the Big-Vul dataset, which contains both vulnerable and fixed versions of C++ functions, along with CVE and CWE identifiers and other related information. Without fine-tuning, ChatGPT lagged behind code-specific models. Beyond GPT's capabilities, Alrashedy et al. [1] examined the CodeLlama model with a Python vulnerability database, using feedback-driven vulnerability fixing, where one instance of the model reviews the response of another, and the answer is then further enhanced based on this review. This approach yielded 5-10% better results than their baseline.

In a more loosely related experiment, Garg et al. [10] used μBERT to mutate existing fixes from Vul4J to artificially simulate vulnerabilities, and then categorized these mutants based on how closely they were coupled to the actual, real-world vulnerabilities by checking whether they fail the same vulnerability-witnessing tests (and whether the fail it for the same reason). They conclude that LLMs are already capable of generating useful synthetic mutants in 39 out of the 45 cases they investigated. A relevant idea to our topic is from Xia et al. [24], which validated the correctness of generated fixes with tests – only here, the tests were preexisting and the fixes were the ones created via model output. In light of the widespread applicability of large language models in this field, we also target the topic of vulnerabilities using GPT-4. However, instead of detecting, synthetically reproducing, or generating fixes for a particular problem, we aim to verify the correctness of existing fixes by generating the corresponding test cases.

## 2.2 LLMs in test case generation

Steenhoek et al. [20] used static metrics to fine-tune models and generate high-quality unit tests. They found that an unrefined model generated syntactically incorrect tests 17% of the time, tests without assertions 31% of the time, and failed to call the method under test 37% of the time. With reinforcement learning from static quality metrics (RLSQM) and supervised fine-tuning, these numbers decreased to around 1%, 5%, and 10%, respectively. They used dynamic context lengths similar to ours, including the entire file's text, a whole class, only the method headers, or the method itself.

Tufano et al. [22] also experimented with different focal context levels during unit test generation. This approach is much closer to the dynamic context we employed. Starting exclusively from the focal method, they incrementally added the class name, constructor headers, method headers, and finally, class fields. The model training was progressive, beginning with English-language pre-training, followed by code-based training, and finally, the fine-tuned model for test case generation. The generation process was interpreted as a translation from the focal context to the test case, and the authors concluded that the more focal context, the better the generation tends to be (with up to 11% improvement).

An empirical study [19] explored the impact of context on test generation, although the context played a different role. In their work, they use different contexts, such as using only the method under test, or the amount of JavaDoc documentation included in the prompt. 40-70% of generated tests did not cause compilation errors, and during subjective evaluation, this proportion was much higher, between 75% and 100%. Generation was correct 50-80% of the time, and code coverage was between 70-90%. Additionally, preliminary work shows that LLMs perform better with weakly typed languages.

A study by Meta [2] focused on improving existing human-written tests rather than generating new tests. 75% of the improvements were syntactically correct, 57% generated successful tests, and 25% improved code coverage. Overall, 10% of the provided test classes were improved, and 73% of the model's recommendations were later accepted by developers.

Mathur et al. [16] used the T5 and GPT-3 models to generate test inputs and/or relevant natural language descriptions. Developers described some preconditions or requirements, and the model's output was the relevant test cases that can be evaluated dynamically.

According to a research survey on software testing with large language models [23], no one has approached model input format the way we have. Typically, only the faulty function (either before or after the fix) or description is provided for generating fixes or tests. However, no one has attempted to pass both the faulty and fixed code to generate test cases. The closest was a study by Kang et al. [13] that used an input format providing a description of the bug for test case generation. Although we also generate test cases, our research does not aim to test a specific functionality for better coverage like the studies mentioned above. Instead, we aim to generate test cases that demonstrate the presence (or absence) of vulnerabilities based on the before/after states of the affected source code, showing failure in the vulnerable state and success in the fixed state.

## 2.3 Test cases for vulnerabilities

The literature most similar to our study considers the intersection of vulnerabilities and test cases. One instance is by Taneja et al. [21], where unit test generation also occurs with the input containing both the old and new versions of the code, but the actual generation is based on instrumentation and path coverage. A more recent example of a similar approach is by Chen et al. [5] where they use genetic test generation to create test cases for library vulnerabilities that have been proven to be reachable from the main application.

Our research also lies within this intersection, but unlike the above, we use an emerging and increasingly promising large language model to generate test cases as evidence of vulnerability instead of using instrumentation and/or genetic algorithms.

## 3 Methodology

Our research process can be divided into three stages: one-time data collection and environment preparation, automatic iteration, and manual validation. A high-level overview of this process is depicted in Figure 1.

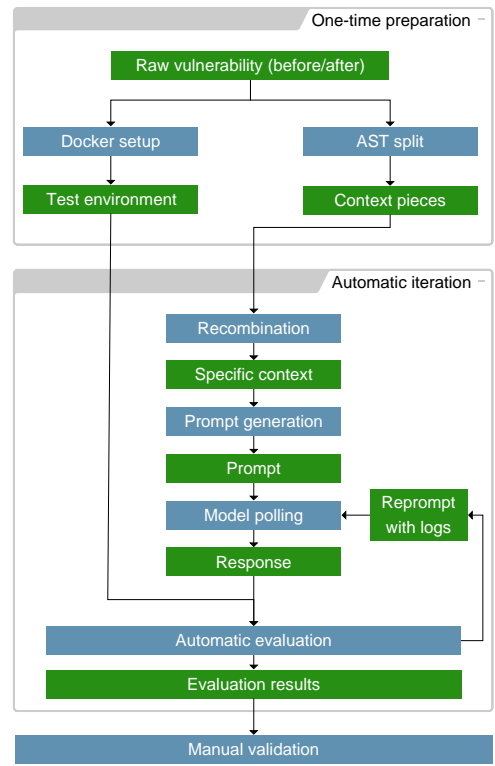

Figure 1: Overview of our methodology

### 3.1 One-time Preparation

To enable automatic evaluation at a later stage, we first needed to collect and standardize the inputs of our framework.

**Data Collection** Our selection criteria were as follows: the dataset should include vulnerable methods, their fixes, and an environment in which these methods could be tested (i.e., the project can be built, and the tests can be executed). Also, all vulnerabilities should exist in real systems instead of being artificial examples.

The VUL4J [4] database, widely used in APR-related research, met these requirements by containing real, certified vulnerabilities from Java projects. This database provides a wide variety of useful information, including vulnerability identifiers, program states before and after fixes, GitHub commit references, and proof of vulnerability (i.e., vulnerability-witnessing) unit tests. Additionally, it includes a Docker environment that significantly simplifies the running of a software's specific version.

The database includes vulnerabilities that can be fixed by modifying one or more methods, potentially involving multiple class and file modifications. To mitigate the risk of the model running out of context, we decided to use only examples where the vulnerability fix affects one single class and, in most cases, one single method. This, combined with the 128k context window of GPT-4 Turbo, prevented any obvious instance of forgetting in our experience.

Within these boundaries, we attempted to include as many different CWE-types of vulnerabilities in our set of 50 subjects as possible to measure GPT-4's unit test generation capabilities across a wide spectrum.

**AST Split** Defining the appropriate amount of contextual information is essential for unit test generation, requiring not only natural language but also the method to be tested. For this purpose, we created context fragments [22] from all vulnerable classes in the collected examples through static analysis and splitting parts of the resulting ASTs (Abstract Syntax Trees) into different bins representing an increasingly large neighborhood of the vulnerable method:

- **L0** includes the class's package declaration, class declaration, and the vulnerable method itself.
- **L1** = **L0** + the headers of the class's constructors, if any.
- **L2** = **L1** + the headers of the class's other methods.
- **L3** = **L2** + the fields declared in the class.

### 3.2 Automatic Iteration

Our automated framework[1] had the most significant role in enabling the scope of this research as it seamlessly handled the context slices and project paths, the communication with the API, the execution of the generated responses, and the evaluation of the results. This process can be summarized in several steps.

**Recombination** To create the textual code snippet for the current round of prompting, we recombined the relevant context slices according to the **L0-L3** guidelines described above. We refer to this snippet as the "focal context". Each of the focal context levels was used in a separate run during the evaluation, so it can be thought of as a separate dimension, whose effects are analyzed in **RQ4**.

**Prompt Generation** As mentioned earlier, generating the unit tests requires a sufficiently large and specific context – and not just the code-based focal context. First, we gathered prompt references

```
You are a senior software tester and a cyber security ↩
    specialist.
You will be given the source code of a Java class where you ↩
    will find the context of a vulnerable method before and ↩
    after the patch.
Your task is to create a unit test that triggers the ↩
    vulnerability and fails before the patch and passes after ↩
    it. The class' name should be the name of the class ↩
    appended with the string "Test".
Use simple Java language features in the generated test!

{focal context of the vulnerable code}

The method after patching the vulnerability:

{patched method of vulnerable code}

It is very important for me, please create the unittest based ↩
    on your best knowledge in the given context.
```

**Listing 1: The final prompt**

from relevant literature [23] and then experimented with the different prompts. Our final version is based on the experiences gained from our initial runs.

According to the official OpenAI documentation[2], specifying a role for GPT helps it better understand the context. In our case, this role is *senior software tester*, which is then included at the beginning of our default prompt setup. Then, we describe the characteristics of the input to be processed and the task the model will work on.

A common error in the generated unit tests is the lack of imports for the functions and classes used. Also, most examples for evaluation have to be compiled with outdated Java versions. This led to the model often generating code that was incompatible with a specific Java language version (i.e., the generated code used features that were introduced in a later version of the language). Therefore, we decided to explicitly warn the model about these details during prompting. The correspondence between the names of classes and their tests also caused problems during automatic evaluation, so we had to define their format in the prompt.

Next came the source code, which we have always added to the prompt as part of the focal context. The vulnerable version of the method first, with context appropriate for the current level (L0-L3), and then the fixed version (without repeating the context to keep the prompt concise and the model's token usage in check).

Finally, we used emotional stimulation, stating the importance of solving the task, which has been shown to yield better results [14]. The final prompt used for all examples in the main evaluation round can be found in Listing 1.

**Model Polling** Once the prompt was complete, polling the GPT-4 API was a practically independent and seamless step. We simply submitted the prompt, parsed the response to extract pure Java source code, then saved this code to a file, and placed it in the project's test environment (in a predefined location per project).

**Automatic Evaluation** The execution of the generated tests was done using the Docker environment provided with the VUL4J database, satisfying all necessary environmental requirements. However, instead of running the whole test bed, the framework ran only the generated test class to reduce runtime.

First, we executed the generated test class on the vulnerable version of the project (expecting it to fail), then the fixed version

---

[1]https://zenodo.org/records/14758148

[2]https://platform.openai.com/docs/guides/gpt/chat-completions-api

(expecting it to pass), storing the execution logs both times. We used these logs to automatically judge the quality of the tests. If it produced the expected fail/pass pattern, it was accepted. Otherwise, we proceeded with reprompting.

**Reprompting** Because of the frequent mistakes in model output, – and because these mistakes could often be attributed to chance as they disappeared when reprompting – we considered performing multiple iterations during the automatic evaluation to avoid random results and examine generation consistency. Instead of raw reiterations, though, we decided to trigger a limited, feedback-based refinement if the previous results did not match the expected pattern (i.e., we either received compilation error(s) or the test results did not match the current code variant's vulnerability).

Since feedback occurs within the same, sufficiently large conversation context, and the model retains previously provided information (thereby making role and technical details redundant to state again), different prompts were associated with these feedback requests. In these, we provided simple, textual feedback to GPT-4 on the violated criterion during test execution, along with the generated log file if necessary. The three situations resulting in a failing response (and their associated prompts) are the following:

- **BEFORE_PASS** - "The test you've provided should have failed for the original version of the vulnerability before the patch, but it passes. Please fix it and return the whole code."
- **AFTER_FAIL** - "The test you've provided should have passed for the patched version of the vulnerability, but it fails. Please fix it and return the whole code."
- **ERROR** - "The code you provided has errors in it: <log>. Fix the error indicated by the compiler message, and answer with the WHOLE fixed code only."

Multiple such feedback iterations were initiated until either three consecutive compilation errors occurred (which we took to mean that the model could not generate a syntactically correct test file in this scenario) or five generations elapsed with no compilation errors but no improvement of the unexpected results either (i.e., the vulnerable code passed the test or the fixed version failed).

**Evaluation Results** Testing was first performed on the vulnerable version (before patch), and then on the fixed version (after patch) of each vulnerability. Each of these two versions had three possible outcomes:

- **PASS**: The generated test ran successfully and passed.
- **FAIL**: The generated test ran successfully but failed.
- **ERR**: The generated test contained compilation errors.

In the end, the best generation was considered for each vulnerability. This means that if a generated test case was compilable at any point, the evaluator retained the resulting log files for that particular test execution, and this became the result we recorded in the final tally (instead of reverting it back to an **ERR** when consequent iterations yielded syntactically incorrect code, for example).

In Section 4, we will discuss some cases where these feedback categories alone did not give the full picture about the results, highlighting the importance of the manual evaluation.

### 3.3 Manual Evaluation

In addition to the fully automated and strict validation, two developers (one of them an author, the other an outside perspective) performed a subjective, manual evaluation of the generated tests as well to look for "usefulness".

Our reasoning for this phase – apart from verifying the validity of the automatic evaluation process – was that even a half-baked test could be better than an empty canvas for a potential test developer if said test was going in the right direction and only failed because of some minor syntax error. After all, developers need extensive, relevant background information to create even a runnable test for a given program. Therefore, we decided to review the structure of each test and manually assess its degree of semantic correctness to evaluate its usefulness, even when the test contained errors. The manual evaluation labeled each result as semantically correct (when both reviewers agreed) or semantically incorrect. Naturally, cases that produced correct outputs during strict validation were also manually checked.

## 4 Results

Table 1 presents the detailed evaluation results: whether the entry passed or failed the final test **Before** and **After** patching, and whether the **Manual** evaluation deemed it "useful". All these for each focal context level (**L0**-**L3**). In total, 200 result pairs were generated during the automated evaluation.

### 4.1 Syntactic analysis

In the syntactic analysis, the question was whether GPT-4 could generate a unit test universally applicable in a test environment without any human intervention, provided that the model was instructed to use simple Java language features. There were instances where the runtime environment would have required extra annotations to be able to execute the test. Since these were missing, some generated tests were not recognized as unit tests by the runtime environment, although they were included in the project's compilation process. Based on this factor, such cases were also accepted as syntactically correct results.

In some runs, where both the vulnerable and fixed versions produced **FAIL** results, the issue was not necessarily with the test result. There were also cases where the compilation did not fail, but the symbols used in the test were unavailable due to missing imports, which the model might have forgotten or which did not exist.

> **Answer to RQ1:** Excluding the manually reviewed cases from the statistics, GPT-4 generated syntactically correct code in 35 out of 50 cases (70%) with **L0** and **L1** focal context. Furthermore, **L2** resulted in 31 correct generations (62%), and **L3** produced positive results in 32 cases (64%), for an overall accuracy of 66.5%.

### 4.2 Semantic Analysis

Of course, syntactic analysis alone is insufficient to determine the generated tests' correctness. In this section, we further filter the syntactically correct cases and examine only the ones that met the automatic validation criteria: the test failed for the pre-fix state (**FAIL**) and succeeded for the post-fix state (**PASS**). Overall, this occurred 15 times (7.5%).

These generated tests always deviated, even if only slightly, from the unit tests included in the VUL4J dataset. To ensure the results' correctness, we also manually evaluated whether they were actually testing what they were meant to – meaning, whether the unit test really verifies the presence of its corresponding vulnerability.

Table 1: The results of the separate runs

| Vul4J ID | L0 | | | L1 | | | L2 | | | L3 | | |
|---|---|---|---|---|---|---|---|---|---|---|---|---|
| | Before | After | Manual | Before | After | Manual | Before | After | Manual | Before | After | Manual |
| VUL4J-01 | PASS | PASS | NO | PASS | PASS | NO | ERR | ERR | OK | PASS | PASS | NO |
| VUL4J-02 | ERR | ERR | NO | FAIL | FAIL | NO | ERR | ERR | OK | ERR | ERR | OK |
| VUL4J-03 | FAIL | FAIL | OK | ERR | ERR | OK | ERR | ERR | OK | FAIL | FAIL | OK |
| VUL4J-04 | ERR | ERR | OK | ERR | ERR | OK | ERR | ERR | OK | ERR | ERR | OK |
| VUL4J-05 | ERR | ERR | OK | ERR | ERR | NO | FAIL | FAIL | OK | ERR | ERR | OK |
| VUL4J-06 | FAIL | FAIL | NO | FAIL | FAIL | OK | FAIL | FAIL | OK | FAIL | FAIL | NO |
| VUL4J-07 | FAIL | FAIL | OK | FAIL | FAIL | OK | FAIL | FAIL | OK | FAIL | FAIL | OK |
| VUL4J-08 | FAIL | FAIL | OK | FAIL | FAIL | OK | FAIL | FAIL | OK | FAIL | FAIL | NO |
| VUL4J-10 | ERR | ERR | NO | FAIL | FAIL | OK | ERR | ERR | NO | FAIL | FAIL | OK |
| VUL4J-12 | PASS | PASS | OK | PASS | PASS | NO | FAIL | FAIL | NO | ERR | ERR | NO |
| VUL4J-13 | FAIL | FAIL | OK | FAIL | FAIL | OK | FAIL | FAIL | OK | FAIL | FAIL | NO |
| VUL4J-16 | FAIL | FAIL | OK | FAIL | FAIL | NO | FAIL | FAIL | OK | FAIL | FAIL | NO |
| VUL4J-17 | FAIL | PASS | OK | FAIL | PASS | OK | FAIL | PASS | OK | FAIL | PASS | OK |
| VUL4J-18 | ERR | ERR | OK | ERR | ERR | OK | ERR | ERR | OK | ERR | ERR | OK |
| VUL4J-19 | FAIL | FAIL | NO | FAIL | FAIL | OK | ERR | ERR | NO | FAIL | FAIL | OK |
| VUL4J-20 | PASS | PASS | NO | ERR | ERR | OK | FAIL | PASS | OK | ERR | ERR | OK |
| VUL4J-22 | FAIL | FAIL | OK | FAIL | FAIL | NO | FAIL | FAIL | OK | FAIL | FAIL | OK |
| VUL4J-24 | FAIL | FAIL | NO | ERR | ERR | NO | PASS | PASS | OK | FAIL | FAIL | OK |
| VUL4J-25 | PASS | PASS | OK | PASS | PASS | NO | FAIL | FAIL | OK | ERR | ERR | OK |
| VUL4J-26 | PASS | PASS | NO | PASS | PASS | NO | ERR | ERR | OK | ERR | ERR | OK |
| VUL4J-30 | PASS | PASS | NO | PASS | PASS | OK | ERR | ERR | OK | PASS | PASS | OK |
| VUL4J-31 | PASS | PASS | OK | PASS | PASS | OK | ERR | ERR | OK | PASS | PASS | OK |
| VUL4J-33 | ERR | ERR | OK | PASS | PASS | NO | ERR | ERR | OK | PASS | PASS | NO |
| VUL4J-34 | PASS | PASS | OK | PASS | PASS | OK | PASS | PASS | OK | PASS | PASS | OK |
| VUL4J-39 | FAIL | PASS | OK | FAIL | FAIL | OK | PASS | PASS | OK | ERR | ERR | OK |
| VUL4J-40 | ERR | ERR | NO | ERR | ERR | OK | FAIL | FAIL | NO | FAIL | FAIL | NO |
| VUL4J-41 | FAIL | FAIL | OK | ERR | ERR | OK | ERR | ERR | OK | FAIL | FAIL | OK |
| VUL4J-43 | PASS | PASS | OK | PASS | PASS | OK | PASS | PASS | OK | PASS | PASS | OK |
| VUL4J-44 | FAIL | FAIL | OK | FAIL | PASS | OK | PASS | PASS | OK | FAIL | PASS | OK |
| VUL4J-45 | FAIL | FAIL | NO | FAIL | FAIL | NO | PASS | PASS | NO | FAIL | FAIL | NO |
| VUL4J-46 | FAIL | PASS | OK | FAIL | PASS | OK | PASS | PASS | OK | FAIL | FAIL | OK |
| VUL4J-47 | ERR | ERR | NO | PASS | PASS | OK | PASS | FAIL | OK | ERR | ERR | NO |
| VUL4J-48 | PASS | PASS | NO | ERR | ERR | OK | PASS | PASS | OK | ERR | ERR | NO |
| VUL4J-50 | ERR | ERR | NO | PASS | PASS | OK | PASS | PASS | OK | PASS | PASS | OK |
| VUL4J-52 | ERR | ERR | OK | ERR | ERR | OK | ERR | ERR | OK | ERR | ERR | OK |
| VUL4J-53 | ERR | ERR | NO | ERR | ERR | NO | ERR | ERR | NO | ERR | ERR | NO |
| VUL4J-54 | ERR | ERR | OK | ERR | ERR | OK | ERR | ERR | OK | ERR | ERR | OK |
| VUL4J-55 | FAIL | FAIL | OK | FAIL | FAIL | OK | FAIL | FAIL | OK | PASS | PASS | OK |
| VUL4J-57 | ERR | ERR | NO | ERR | ERR | NO | ERR | ERR | OK | ERR | ERR | NO |
| VUL4J-60 | FAIL | FAIL | OK | FAIL | FAIL | OK | ERR | ERR | NO | PASS | PASS | OK |
| VUL4J-61 | PASS | PASS | NO | ERR | ERR | OK | PASS | PASS | NO | PASS | PASS | OK |
| VUL4J-62 | PASS | PASS | NO | PASS | PASS | NO | PASS | PASS | NO | PASS | PASS | OK |
| VUL4J-63 | FAIL | PASS | OK | FAIL | PASS | OK | FAIL | FAIL | OK | ERR | ERR | OK |
| VUL4J-66 | FAIL | FAIL | NO | FAIL | FAIL | OK | FAIL | FAIL | OK | FAIL | FAIL | OK |
| VUL4J-69 | FAIL | FAIL | NO | FAIL | FAIL | NO | FAIL | FAIL | OK | FAIL | FAIL | OK |
| VUL4J-73 | FAIL | FAIL | OK | FAIL | FAIL | OK | FAIL | FAIL | OK | FAIL | FAIL | OK |
| VUL4J-74 | FAIL | FAIL | NO | FAIL | FAIL | NO | FAIL | FAIL | NO | FAIL | FAIL | OK |
| VUL4J-75 | ERR | ERR | OK | PASS | PASS | OK | ERR | ERR | OK | PASS | PASS | NO |
| VUL4J-76 | PASS | PASS | NO | FAIL | FAIL | OK | PASS | PASS | OK | FAIL | FAIL | OK |
| VUL4J-77 | FAIL | PASS | OK | FAIL | PASS | OK | FAIL | PASS | OK | ERR | ERR | NO |

Of the remaining 92.5%, the following common failure patterns were observed – which might help with future prompt design:

- 55 import errors, where the fix is simply importing some missing symbols,
- 33 undetected tests, where our framework did not execute the tests automatically,
- 20 mistargeted tests, where the method under test would have been redefined or substituted,
- 8 misused tests, where the correct method is called, but the call signature is incorrect,
- 5 visibility errors, where the test didn't have permission to call a protected or private method,
- 3 version errors, where the generated test uses language features inconsistent with the subject system,
- 61 various "other" errors, including dependency errors, real semantic compile errors, etc.

We do note, however, that this 7.5% figure can be misleading when also considering the subjective usefulness of the generated tests – which we look at next.

**Answer to RQ2:** A total of 15 (7.5%) generated tests were not only syntactically, but also semantically correct. According to focal context levels, **L0**, and **L1** produced 5 correct results (10%), while **L2** produced 3 (6%), and **L3** only 2 (4%) positive results in an interesting reversal of the expected context effect.

## 4.3 Subjective usability of the generated tests

To assess the subjective usability of the generated test cases, two developers independently analyzed all final results from the automatic phase and determined the amount of effort needed to fix them. The main evaluation criteria for labeling a test "useful" were the following:

- The generated test is relevant to the particular entry, i.e., the vulnerable code and its fix.
- The generated test actually tests the target method.

```
@Test
public void testMethod() throws Exception {
  ...
  JpegDecoder jpegDecoder = new JpegDecoder();
  // Call the private extend method using reflection
  int result = (Integer) extendMethod.invoke(jpegDecoder, v, t);
  // Hypothetical patched logic that accounts for integer ↩
      overflow
  long expected = ((long) v) + ((v < (1 << (t - 1))) ? ((-1L) <<↩
       t) + 1 : 0);
  int expectedInt = (int) expected;
  assertEquals("The extend method should behave as expected ↩
      after the patch", expectedInt, result);
}
```

**Listing 2: Details of the L1 VUL4J-12 test case**

```
...
int vt = (1 << (t - 1));
- while (v < vt) {
+ if (v < vt) {
  vt = (-1 << t) + 1;
  v += vt;
}
...
```

**Listing 3: The fix for the VUL4J-12**

- The generated test can be corrected with minimal human refinement (e.g., handling missing dependencies, unsupported language features, or incorrect call signatures).
- The generated test does not attempt to override the original method under test.

We argue that if a test fulfills all these requirements, then it is already better than an empty canvas when trying to test its corresponding method. So, it does have value, even if it happens to be syntactically incorrect yet, and needs a bit of manual intervention.

An example of a subjectively rejected test is the one generated for *VUL4J-24* during the **L0** run. In this case, the model tried to override the behavior of the method under test instead of correcting the test based on the provided context. Feedback-based generation also played a role in this phenomenon.

Additionally, in some cases, the model had difficulty interpreting the provided source code. In the **L1** generation, the provided code for *VUL4J-12* was misunderstood. *VUL4J-12*'s fix involved eliminating an infinite loop (as can be seen in Listing 3), but the test attempted to compare a *long* value to an integer using type casting (shown in Listing 2). Although these were not random instructions, as they were relevant to the provided source code, the generated test did not adequately test the presence of a vulnerability; thus, we rejected the test in manual evaluation.

On the other hand, many useful tests were generated. A great example is the test for *VUL4J-5* in **L3**, presented in Listing 4.

Although the generated test was syntactically incorrect, the model correctly interpreted the task and the generated test actually tested the presence of the vulnerability. In the vulnerable code, exploiting a path traversal vulnerability was possible by passing the input `"../file"`. In this case, the path can point to anywhere on the file system (even outside the working directory), thus endangering many files that should not be accessible externally. The test's mistake was an incorrect call signature to the relevant expand

```
@Test
public void testExpandVulnerability() {
  ...
  // Vulnerable entry trying to write outside of target ↩
      directory
  ArchiveEntry vulnerableEntry = new ArchiveEntry("../↩
      outsidetarget.txt", false);
  ...
  try {
      // This should fail before patch and pass after patch
      expander.expand(supplier, writer, targetDirectory);
      fail("The test should throw an IOException before the ↩
      patch as it attempts to expand outside of the target ↩
      directory.");
  } catch (IOException e) {
      // Expected exception
      String expectedMessage = "expanding ../outsidetarget.txt ↩
      would create file outside of " + targetDirectory;
      assertEquals(expectedMessage, e.getMessage());
  }
}
```

**Listing 4: Relevant parts of the generated test case (VUL4J-5, L3)**

```
@Test
public void test_withUnsafeFinalizeMethod() {
  thrown.expect(SecurityException.class);

  MethodNode unsafeFinalizeMethod = new MethodNode(
        "finalize",
        Opcodes.ACC_PUBLIC,
        ClassHelper.VOID_TYPE,
        Parameter.EMPTY_ARRAY,
        ClassNode.EMPTY_ARRAY,
        null
  );

  classNode.addMethod(unsafeFinalizeMethod);
  transformer.transform(sourceUnit, null, classNode);
}
```

**Listing 5: The L2 VUL4J-52 test case**

method (which is a one-liner fix), and a few unnecessarily generated mock inner classes (that can be deleted after the signature fix, but would not hurt even if they stayed).

Another example (which can be seen in Listing 5) is the test generated for *VUL4J-52* at the **L2** context level. To fix an *improper access control* vulnerability, a method was implemented to prohibit the override of the *finalize* method in the sandbox environment, accompanied by throwing a *SecurityException*.

The generated test correctly checks the presence of such a function and expects the appropriate exception, which would not occur in the vulnerable code but occurs in the fixed version. Its only flaw was the usual missing imports that prevented it from being automatically flagged as a **PASS**.

> **Answer to RQ3:** Based on the criteria mentioned above, 68.5% of the generated tests could be considered useful. These tests provide a usable skeleton for the developers to create appropriate vulnerability-witnessing tests with only minor modifications.

### 4.4 Impact of context on accuracy

Although the previously presented results did not show significant syntactical or automatically evaluable differences across different context levels, some manual observations were made during the research. We performed this added investigation as, for example, the context levels in the work of Michele et al.[22] positively impacted

```
@Before
public void setUp() {
  factory = new DiskFileItemFactory();
}

@Test(expected = IOException.class)
public void testReadObjectWithInvalidRepositoryPath() throws ↩
    Exception {
  [...]
  File invalidRepository = new File("/some/repository\0");
  diskFileItem.setRepository(invalidRepository);
  [...]
```

**Listing 6: Example code from the L0 run (excerpt)**

```
@Before
public void setUp() {
  // Set up a valid repository directory for DiskFileItem
  repository = new File(TEMP_DIR, "diskFileItemTest");
  repository.mkdir();

  diskFileItem = new DiskFileItem(FIELD_NAME, CONTENT_TYPE, ↩
      IS_FORM_FIELD, FILE_NAME, SIZE_THRESHOLD, repository);
}
```

**Listing 7: Example code from the L1 run (excerpt)**

the results in nearly 40% of cases using a BART-based model with code-based pre-training and fine-tuning for test case generation.

Our manual evaluation indicated that the model does write more deliberate tests with fewer errors when aware of the constructors' structure. For instance, when using the **L0** context level, GPT-4 confused the publicly available Java *FileItem* class with the *DiskFileItem*, which is a class defined in the project. When the constructor's prototype was introduced in **L1**, the model generated a semantically correct test. The example of the resulting test codes are shown in Listing 6 and Listing 7 for the runs **L0** and **L1**, respectively.

Based on the results discussed in Section 4.1, the best-performing context level was **L0** and **L1**, both with 35 correct generations, and additional context only seemed to confuse the model, resulting in declining percentages. From the perspective of subjective usability, though, **L2** was the winner, producing 40 "useful" generations. So despite the observation that extra context did not seem to help (or even hurt) syntactically, or in the automatic evaluation of semantics, it did show a definite positive effect in the subjective usability of the generated test cases. This observation is much closer to the original results by Tufano et al. [22] as well, who concluded that "the more context, the better" – even though **L3** exhibited a slight decline for us.

> **Answer to RQ4:** Based on the subjective evaluations, extra context helps the model solve tasks more precisely. Using the **L0** context level, 28 cases (56%) were considered semantically useful, 34 cases (68%) for **L1**, 40 cases (80%) for **L2**, and 35 cases (70%) for **L3**. Our results show that adding a certain amount of code context leads to significant improvements.

## 5 Discussion

### 5.1 Feedback-based generation observations

Feedback-based generation was a fundamental principle of our approach. This significantly aided correct code generation in many

```
It's important to note that with the SafeConstructor, the YAML ↩
    parser should not throw a SecurityException but rather a ↩
    YAMLException, ConstructorException, or another related ↩
    exception when encountering malicious YAML content...
@Test
public void testReadYamlTreeVulnerability() {
  ...
  assertTrue("Expected a YAML parser exception due to patched ↩
    vulnerability",
              e instanceof org.yaml.snakeyaml.error.↩
    YAMLException || // change this if needed
              e instanceof org.yaml.snakeyaml.constructor.↩
    ConstructorException); // change this if needed
}
```

**Listing 8: Generated response for VUL4J-77 L2 run (excerpt)**

situations, as the model interactively retrieved log file content related to its generated code. For example, the generated code for the *VUL4J-77* instance at **L2** context level initially returned a **FAIL** - **FAIL** result pair, which triggered the feedback-based generation. GPT-4 refined the code based on the error log the framework provided, eventually producing completely valid results (i.e., a test case that yielded **FAIL** - **PASS** for the original and fixed code, respectively). An excerpt from the response is shown in Listing 8.

However, in some cases, the model completely "lost the plot" and took test code refinement in a direction irrelevant to fixing the provided issue(s). The initial test code created for the *VUL4J-62* example in **L1** context did not yet include task-irrelevant code generation. It is important to note that for some test runs, the test class had to extend the TestCase [3] class; otherwise, the plugin used in such projects did not recognize the test class. This situation was also present in this example, meaning the log files only indicated that no tests were run and the build process was successful. These cases always formed a **PASS** - **PASS** result pair, triggering the feedback-based generation using the **BEFORE_PASS** prompt. In these cases, GPT-4 only guessed and never managed to resolve the issue.

### 5.2 Prompt ablation

To check how much each of the "additional" prompt elements on top of the before-after code states contributed to the final accuracy, we also performed an ablation study on including the role, the emotional appeal and the CWE identifier. Besides the **baseline** setup (i.e., with role and emotional appeal but *without* CWE identifiers), we completed three more reruns of our entire experiment toolchain with the following configurations:

- **no_emotion**: The same prompt as the baseline, only without the emotional appeal of "It is very important for me, please create the unittest based on your best knowledge in the given context" part at the end.
- **no_role**: A further restriction of no_emotion, where not only the emotional appeal, but also the role specification is omitted from the prompt (i.e., the "You are a senior software tester and a cyber security specialist" part).
- **with_cwe**: An expanded version of the baseline prompt with additional CWE info accompanying the vulnerability.

The effect of these configurations is summarized in Table 2.

---
[3] JUnit 4 provided the opportunity to extend TestCase class

Table 2: The results of the prompt ablation

| Config      | Syntax | Semantics |
|-------------|--------|-----------|
| `baseline`  | 66.5%  | 7.5%      |
| `no_emotion`| 68.5%  | 4.5%      |
| `no_role`   | 62.5%  | 4.5%      |
| `with_cwe`  | 65.0%  | 4.0%      |

From a syntactic perspective, only the `no_emotion` prompt helped slightly compared to the baseline, with the other two variants decreasing the percentage of correct generations. The `no_role` configuration was the worst, indicating that setting the appropriate frame for the model to operate in early on is indeed a low hanging fruit. Interestingly, both `no_emotion` and `with_cwe` behaved contrary to expectations – with removing emotional appeal *improving*, while adding extra vulnerability categorization *decreasing* syntactic correctness by 2% and 1.5%, respectively.

From the perspective of semantics, all three alternatives deteriorated generation quality. Missing either emotion only, or both emotion and a role specification, lead to a 3% decrease. In another surprise observation, adding CWE labels hurt the process even more than omitting either emotion or roles with a 3.5% decrease.

While these results are far from generalizable – and the effect of the LLM's creativity and temperature settings have not been isolated yet – it is already safe to say that the contents and structure of the prompt have a measurable effect on the quality of test case generation.

### 5.3 CWE-based trends

Another interesting aspect we investigated is what effect the CWE-grouping of a given vulnerability has on the performance of the test case generation, if any. We only considered CWE IDs that had at least four different vulnerabilities mapping to them to prevent individual outliers from skewing the statistics. Additionally, a special category of "Not Mapping" is also present for the vulnerabilites that did not have corresponding CWE information in the source Vul4J database. We then compared these subsets' syntactic correctness, automatic semantic evaluability, and subjective usefulness to the overall average. Our results are presented in Table 3.

Table 3: The results of the CWE grouping

| Subset      | Syntax |        | Semantics |        | Usability |        |
|-------------|--------|--------|-----------|--------|-----------|--------|
| Average     | 66.5%  | –      | 7.5%      | –      | 68.5%     | –      |
| CWE-20      | 66.7%  | +0.2%  | 0.0%      | -7.5%  | 54.2%     | -14.3% |
| CWE-22      | 70.0%  | +3.5%  | 0.0%      | -7.5%  | 85.0%     | +16.5% |
| CWE-79      | 81.3%  | +14.7% | 0.0%      | -7.5%  | 81.3%     | +12.7% |
| CWE-611     | 56.3%  | -10.3% | 0.0%      | -7.5%  | 50.0%     | -18.5% |
| CWE-835     | 82.1%  | +15.6% | 0.0%      | -7.5%  | 60.7%     | -7.8%  |
| Not Mapping | 61.1%  | -5.4%  | 25.0%     | +17.5% | 88.9%     | +20.4% |

According to these results (and their differences from the overall average), CWE-20 and CWE-611 seem like harder tasks for automatic generation. For CWE-20 (Improper Input Validation), this is reflected in usability only, while for CWE-611 (External XML Entity References), even syntax is affected. On the other hand, CWE-22 (Path Traversal) and CWE-79 (Cross-site Scripting) seem easier to test than usual – the former with slight syntactic and considerable usability impact, the latter with meaningful increases on both fronts. An in-between category is CWE-835 (Infinite Loop), which appears the easiest to generate syntactically correct tests for, yet harder to generate something actually useful for.

From the perspective of automatically evaluable semantic correctness, everything except the "Not Mapping" category scored 0%. This means that every vulnerability that was fully and flawlessly solved without human interaction in our study is either not mapped to a CWE-group, or is a one-off from an underrepresented CWE-group where we do not have enough data for even rudimentary generalization. The unmapped category, apart from being best in semantics, is also the best in subjective usability by producing useful test in 88.9% of cases – yet it is somewhat worse than usual when pure syntax is concerned, with 5.4% below the average. This area definitely warrants further investigation in the future.

## 6 Threats to Validity

We examined only GPT-4's unit test generation capabilities. Future research should investigate other models, particularly code-specialized ones such as CodeLlama. We did not employ any pre-training or fine-tuning. To avoid the inconsistencies we mentioned earlier, it might be worthwhile to follow Michele et al. [22] and subject some models to code-based pre-training followed by test generation fine-tuning. On the other hand, our results might better reflect the "off-the-shelf" capabilities of GPT-4 this way.

The prompts we used were structurally well-organized, and we did try different versions regarding the additional techniques included, but all of our prompts were essentially variations on a single theme. We didn't include additional project data, try different roles/emotions, or experiment with advanced strategies like Chain-of-Thought or Retrieval Augmented Generation. Future efforts should focus more on prompt engineering since it measurably affects generation accuracy.

We used VUL4J [4], a manually assembled collection of vulnerabilities, meaning it may not fully represent the entire spectrum of vulnerabilities, potentially limiting how well our findings generalize to other domains or languages. Future research should use larger, more diverse datasets, which our methodology may help make more accessible. Although VUL4J was released before the GPT training cut-off, data leakage should not affect our results [18].

The subjective usability assessment might be influenced by the experience of the individuals performing the evaluation. To mitigate this limitation, evaluators reviewed and analyzed each other's work and the associated comments for each example. We maintain, however, that fixing "almost correct" code is easier than writing fully correct code from scratch. And while "almost correct" is harder to formalize, future research should consider a more granular approach to test generation than the binary correct/incorrect.

GPT-4 is non-deterministic, meaning that during the production of results, we never received the same answer twice for a given example. This behavior poses a challenge in replicating results exactly and assessing the model's consistency. We did not specify temperature settings during prompting, though Guilherme et al. [11] suggest these could influence test case generation. Experimenting with different temperatures could be useful in the future.

## 7 Conclusion

Our research demonstrates the unit test generation capabilities of GPT-4 within the context of real-life vulnerabilities. Unlike previous studies, our approach uses vulnerable and fixed methods to aid test code generation. In summary, GPT-4's unit test generation capabilities, while far from perfect, do show promise (even without fine-tuning). When a generated test code is not syntactically correct, there is still a good chance that minimal human effort can produce a correct test case from it. We showed that different levels of context affect the generation quality. We experienced that re-prompting can be worthwhile. Feedback that contains sufficient information for correcting an incorrect test case generally pushes the code generation in the right direction. Based on our results, a fully automated solution is not yet feasible, as a 7.5% chance at a seamlessly working solution is just not high enough to warrant adoption. We can say, however, that with a usefulness rate of 68.5%, generating vulnerability-witnessing unit tests with large language models can already provide helpful assistance for developers, as well as researchers trying to expand existing vulnerability databases. And with the many potential improvement directions available, we are confident that further research is only going to increase LLMs' performance in this field.

## Acknowledgments

This work was supported in part by the European Union project RRF-2.3.1-21-2022-00004 within the framework of the Artificial Intelligence National Laboratory; and in part by the Project no TKP2021-NVA-09 implemented with the support provided by the Ministry of Culture and Innovation of Hungary from the National Research, Development and Innovation Fund, financed under the TKP2021-NVA funding scheme. The work has also received support from the European Union Horizon Program under the grant number 101120393 (Sec4AI4Sec).